\documentclass{article}
\usepackage[english]{babel}
\usepackage[utf8]{inputenc}
\usepackage{bm}
\usepackage{amsmath,amsthm,amsfonts}
\usepackage{graphicx}
\usepackage{color}
\usepackage{multirow}
\usepackage{subcaption}
\usepackage{fancyhdr}  
\usepackage{geometry}  
\usepackage{caption}

\graphicspath{ {./figs/} }

\date{}

\newcommand{\mytitle}{A Simulation Method to Play with Porosity and Particle Size Distribution for Characterizing Porous Media}

\title{\mytitle \thanks{This note is based on the author's patent CN112461718B.}}

\author{
Yuhe Wang \thanks{National \& Local Joint Engineering Laboratory for Big Data Analysis and Computing Technology, Beijing 100190, China.}
\thanks{Institute for Scientific Computation, Texas A\&M University, College Station, Texas 77843, USA. Email: {\tt yuhe.wang@tamu.edu}.}
}

\date{\normalsize \today}

\begin{document}

\maketitle

\thispagestyle{empty}

\pagestyle{fancy}  

\begin{abstract}
This note presents a simulation method for investigating the relationship between porosity and particle size distribution in porous media characterization. The method simulates particle packing based on particle size distributions, incorporating descent and collision rules to mimic the natural deposition process while minimizing boundary-induced porosity distortions. By systematically varying the mean particle size and standard deviation, multiple packed structures are generated, and their corresponding porosities are calculated. This approach offers a flexible and scalable tool for exploring the impact of particle size distribution on porosity, providing both qualitative and quantitative insights into porous media properties. 
\end{abstract}

\section{Background}

Porosity is a critical parameter in understanding the behavior of porous media and plays a vital role in various geological and engineering processes, including petroleum recovery \cite{yan2018enhanced,wang2020generalized, jin2015consideration}, carbon dioxide sequestration \cite{pu2018co2, medina2017characterization, tripathy2018comparative}, hydrogen storage \cite{song2023underground, lysyy2022pore, ren2014shaping}, groundwater flow \cite{mi2017enhanced, li2019particle, urumovic2016referential}, and the transport of contaminants \cite{moutsopoulos2001hydraulic, el2005multiple, shapiro2017porosity} in subsurface environments. Accurate characterization of porosity is essential for understanding fluid flow behavior in natural porous systems \cite{du2020connectivity, anovitz2015characterization, yan2016beyond}. Traditional methods for measuring porosity—such as mercury intrusion porosimetry, fluid displacement techniques, and sieving—are well-established but are often time-consuming, labor-intensive, and can require sophisticated laboratory setups \cite{danielson1986porosity,espinal2002porosity, saidian2014comparative}. Given the limitations of these experimental approaches, there has been increasing interest in computational modeling and simulations as more efficient alternatives for understanding pore structures and their relationship to particle size distribution \cite{mcgeary1961mechanical, wu2018reconstruction, zhang2018new, sun2018structural, sadeghnejad2020multiscale, fadlelmula2016separable}.

Recent advancements in high-resolution digital imaging techniques have allowed direct observation of the pore structures within rock core samples. Techniques such as X-ray computed tomography (CT) and scanning electron microscopy (SEM) offer detailed insights into pore geometry at the microscale \cite{akin2003computed, ni2021quantitative,hemes2015multi}. However, due to the inherent complexity of pore structures in natural rock samples, calculating porosity directly from imaging data is challenging \cite{callow2020optimal, elkhoury2019resolution}. Natural porous media exhibit significant heterogeneity, with irregular pore shapes and distributions that complicate traditional measurement techniques. Computational simulations have emerged as a powerful tool to overcome these challenges by modeling the process of particle deposition and packing, allowing for the dynamic visualization of pore formation and facilitating more detailed investigations of porosity and its controlling factors \cite{lee2021computational, elimelech2013particle}.

\begin{figure}
    \centering
    \includegraphics[width=0.5\linewidth]{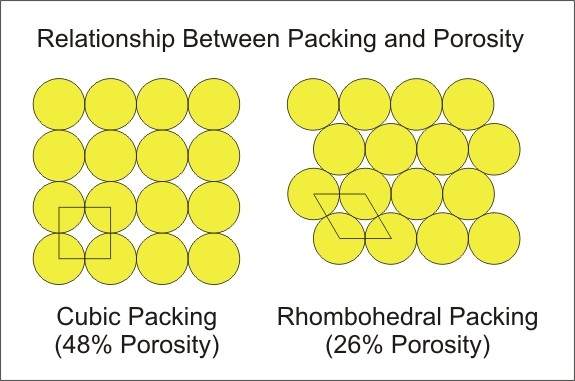}
    \caption{Variations in porosity as a function of cubic and rhombohedral packing of sand grains. Image from \cite{lake2012reservoir}}
    \label{fig:packing}
\end{figure}

Porosity is known to depend on particle size distribution \cite{peronius1985correlation}, but a comprehensive quantitative understanding of this relationship remains elusive. Numerous studies have demonstrated that particle size influences the arrangement and packing of particles, which in turn affects the resulting pore space and overall porosity of the medium. For example, in systems composed of uniform spherical particles, different packing configurations lead to significantly different porosities (\textbf{Figure \ref{fig:packing}}): cubic packing yields a porosity of 47.6\%, while rhombohedral packing result in porosities of 26\%, respectively \cite{hook2003introduction, pal2020influence, lake2012reservoir}. While such models provide useful insights, they are overly simplistic for characterizing natural rock formations, where particle size distributions are non-uniform, and packing arrangements are considerably more complex.

Previous research has focused on the role of average particle size in determining porosity, as it provides a simple and convenient metric for describing the system. However, average particle size alone is insufficient to capture the full complexity. Particle size distribution, which includes both the mean particle size and the variance or standard deviation of particle sizes, provides a more comprehensive description of the system \cite{peronius1985correlation, taleghani2017study}. The standard deviation in particle sizes is particularly important, as it reflects the degree of heterogeneity within the sample. It is commonly understood that large particles can have two opposing effects on porosity. On one hand, they occupy the space that smaller particles and pores could fill, reducing the overall porosity. On the other hand, large particles may prevent smaller particles from packing tightly, potentially increasing porosity. Some researchers believe that both effects are present during the particle packing process, but the degree to which each influences the final porosity varies. Despite this understanding, the quantitative relationship between particle size distribution and porosity has yet to be fully clarified \cite{white1937particle, latham2002prediction}.

The particle-based simulations can model the dynamic process of particle accumulation and packing \cite{potyondy1996modelling, potyondy2004bonded}. These simulations are valuable because they not only reduce the time required for porosity analysis but also allow for the visualization of complex packing processes, which are difficult to capture using traditional experimental techniques \cite{latham2002prediction, yang2000computer}. Despite significant progress in understanding the qualitative relationship between particle size distribution and porosity, a robust quantitative model is still needed. A key reason for this gap is that most studies have focused on idealized systems with uniform particle sizes or simplified packing arrangements, whereas natural systems are much more complex. In natural rock formations, particles exhibit a range of sizes, shapes, and degrees of sorting, all of which influence porosity in ways that are not fully captured by existing models. Moreover, the interactions between particles of different sizes can lead to the counterintuitive effects mentioned above.

In this note, a two-dimensional particle packing simulation method is presented to characterizes the relationship between porosity and particle size distribution. The approach focuses on simulating the natural deposition process of rock particles and investigating how different size distributions affect the resulting pore structure. By using this simulation-based method, the aim is to provide a more detailed and accurate description of the relationship between particle size distribution and porosity. This method can addresses the limitations of traditional experimental techniques. In addition, it offers a scalable solution for better understanding the important porous system.

\section{Method}

In this note, it is focus on a two-dimensional scenario of an efficient two-dimensional particle packing simulation method to characterize the relationship between porosity and particle size distribution. The three-dimensional extension can be further derived base on the method described here. A rectangular simulation domain is used to represent a two-dimensional plane, with dimensions defined by length $L$ and height $H$. For simplicity, dimensionless units are adopted, allowing the model to be applied across a wide range of systems without the need for specific physical units.

Particles are randomly generated along the horizontal axis within the domain, with their positions uniformly distributed between \(0\) and \(L\). For simplicity, it is assumed that all generated particles fall from the same vertical height \(y\), which is sufficiently large (\(y \gg H\)) to ensure uniformity in the initial conditions for particle packing. The height is sufficiently large, so the initial height of each particle can be considered the same. The randomness of the particle locations reflects natural stochasticity in particle deposition processes.

In addition to spatial location randomness in the horizontal axis, the size of the particles plays a crucial role in determining the final packing arrangement and overall porosity. To accurately model these effects, the particle size distribution must be carefully defined. The distribution is characterized by the mean particle radius, \(\mu\), and the standard deviation, \(\sigma\). If the particles follow the normal distribution, then the Probability Density Function (PDF) can be written as:

\begin{equation}
    P(r) = \frac{1}{\sigma \sqrt{2\pi}} e^{-\frac{(r - \mu)^2}{2\sigma^2}}
\end{equation}

where $r$ is the particle radius, $\mu$ is the mean particle radius, and $\sigma$ is the standard deviation. 

The corresponding Cumulative Distribution Function (CDF) is then:

\begin{equation}
    C(r) = \frac{1}{2} [1 + \textit{erf}(\frac{r-\mu}{\sigma \sqrt{2}})]
\end{equation}

where \textit{erf} is the error function, \(\textit{erf}=\frac{2}{\sqrt{\pi}}\int_{0}^r e^{-t^2}dt \). 

However, real-world systems, especially geological formations, often have a more complex distribution of particle sizes that may not be fully captured by a standard normal distribution \cite{blott2012particle}. To address this, a weighted correction is applied to the normal distribution to better represent the proportions of small and large particles. Using the inverse square of the particle radius $r$, the weighted corrected PDF is: 

\begin{equation}
\label{correctedPDF}
    \hat{P}(r) = \frac{\frac{P(r)}{r^2}}{\int_{\mu-2\sigma}^{\mu+2\sigma}\frac{P(r)}{r^2}}             
\end{equation}

The weighted corrected PDF (\textbf{Equation \eqref{correctedPDF}}) is then integrated to form the weighted corrected $CDF$ (\textbf{Equation \eqref{correctedCDF}}), which is to be used to generate random particle sizes. The weighted corrected $CDF$ is essential for generating realistic particle sizes that reflect the observed ratio of small to large particles in natural systems. 

\begin{equation}
\label{correctedCDF}
    \hat{C}(r) = \frac{\int_{\mu-2\sigma}^r \hat{P}(r) dr}{\int_{\mu-2\sigma}^{\mu+2\sigma} \hat{P}(r) dr}
\end{equation}

Once the particle size distribution is established, particle radii are sampled from the weighted corrected $CDF$, ensuing that each generated particle size adhere to the realistic distribution. Each particle is assigned a random horizontal coordinate $x$ within the simulation domain. As mentioned above, the vertical coordinate $y$ is set at a high value above the simulation domain, ensuring that all particles fall from the same height. 

\begin{figure}[h]
    \centering
    \includegraphics[width=0.3\linewidth]{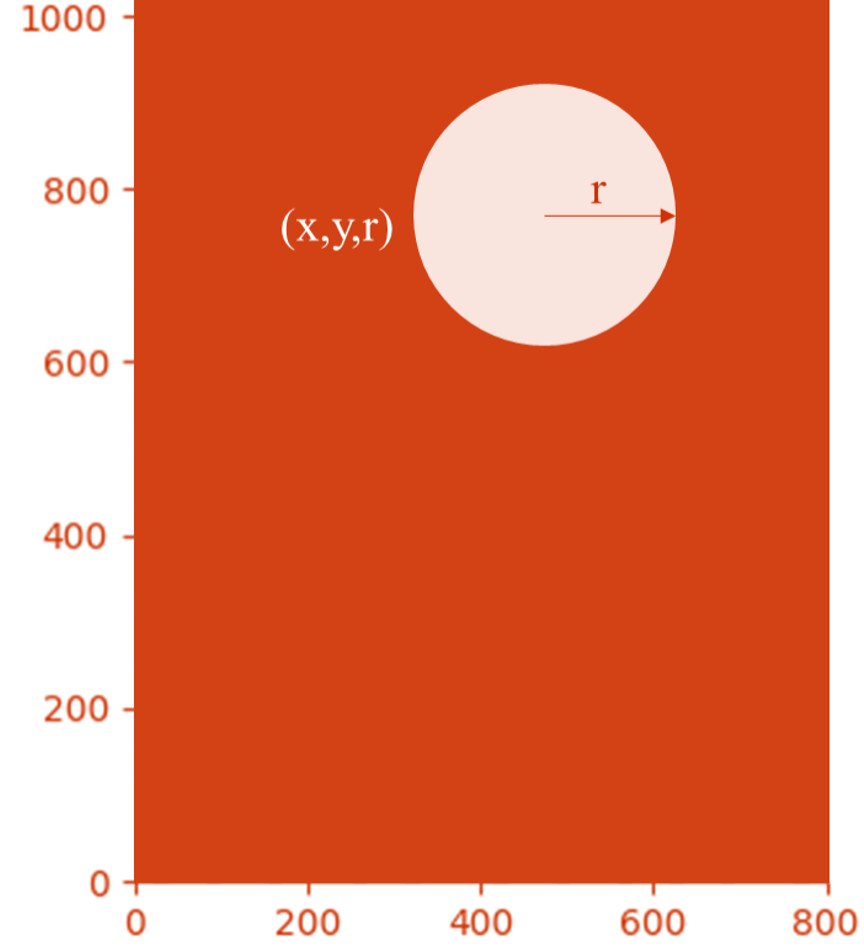}
    \caption{New particle generated using the weighted corrected CDF}
    \label{fig:newparticle}
\end{figure}

A random probability value is sampled from a uniform distribution over the range \( (0,1) \), ensuring equal likelihood for every possible value. This probability is then applied to the weighted corrected cumulative distribution function \( \hat{C}(r) \). By inverting \( \hat{C}(r) \), a corresponding particle radius \( r \) is obtained. Using this radius \( r \) and a randomly generated horizontal coordinate \( x \), a circle with center \( (x, y) \) and radius \( r \) is defined, representing a two-dimensional particle in the simulation (\textbf{Figure \ref{fig:newparticle}}). The vertical coordinate \( y \) is set to a value greater than the height of the simulation area, ensuring that the vertical position does not interfere with the particle packing process.

\begin{figure}[h]
    \centering
    \includegraphics[width=0.3\linewidth]{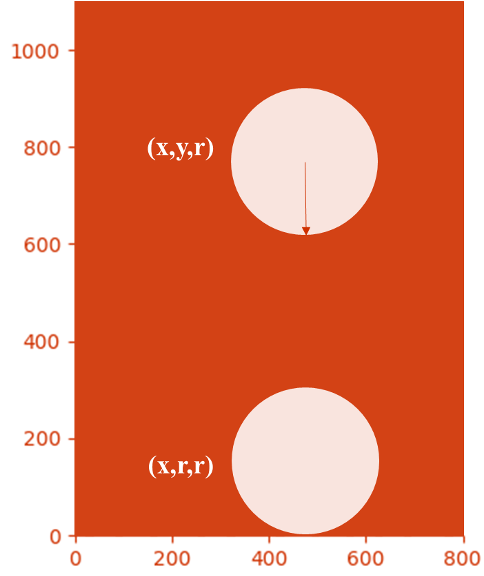}
    \caption{Descending new particle: no collision}
    \label{fig:noncollision}
\end{figure}

As particles descend, the simulation models their interactions and collisions with previously placed particles based on certain descending and collision rules as follows. 
If a particle has no other particles below it, it settles on the bottom boundary of the simulation domain. The final position of this particle is \((x,r,r)\), as illustrated in \textbf{Figure \ref{fig:noncollision}}.

\begin{figure}[h]
    \centering
    \includegraphics[width=0.3\linewidth]{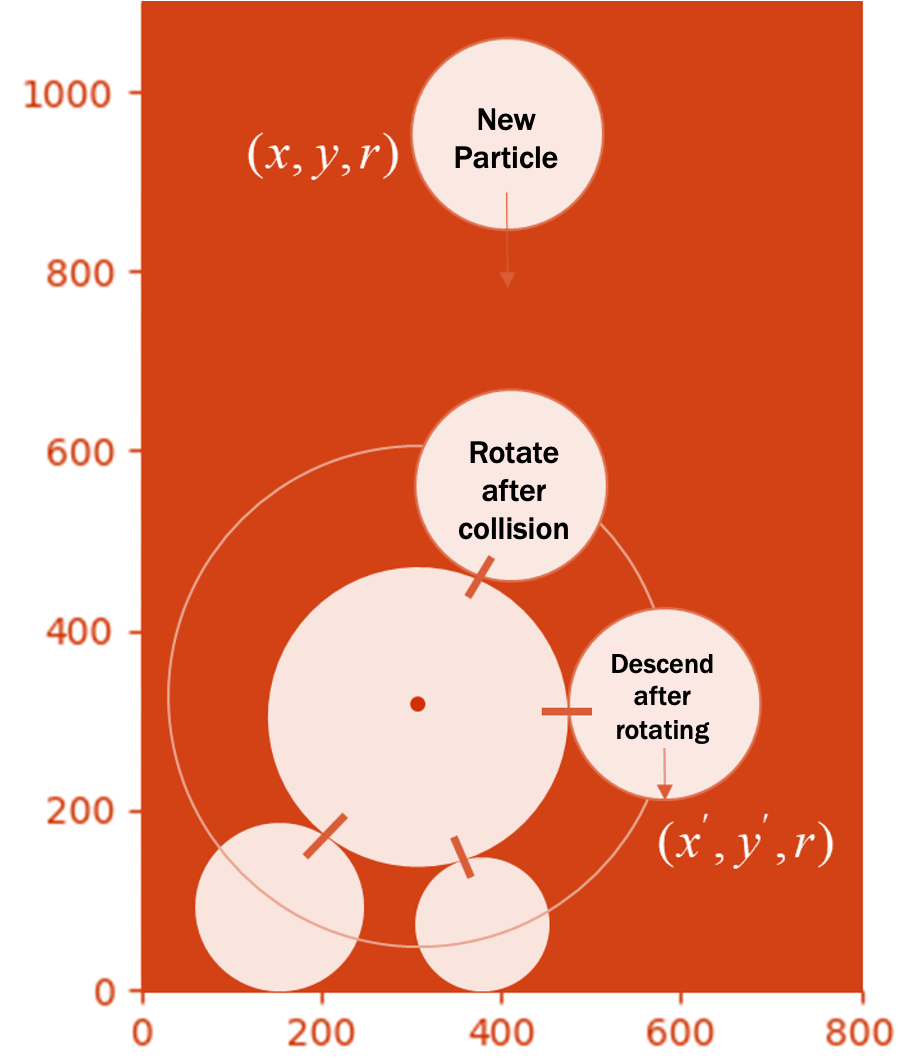}
    \caption{Descending new particle: collision and rotation}
    \label{fig:collision1}
\end{figure}

If there are already particles in the space below, the new particle collides with them. In this case, the particle may rotate or shift until it comes to rest in a stable position atop the previously settled particles. The process assumes infinite friction between particles, which prevents the particles from sliding after the collision. The new particle \( (x, y, r) \) rotates upon colliding with the existing particle \( (x_1, y_1, r_1) \), moving until it reaches the same height as the center of the colliding particle, such that \( y = y_1 \). After the new particle \( (x, y, r) \) moves to the new position \( (x_2, y_1, r) \), it continues its descent. The process of descent and collision is repeated, continuously updating the position of the newly generated particle until it either reaches the ground or stabilizes by resting on two already existing particles. At this point, the descent and collision process for the particle is considered complete (\textbf{Figure \ref{fig:collision1}}).

\begin{figure}
    \centering
    \includegraphics[width=0.3\linewidth]{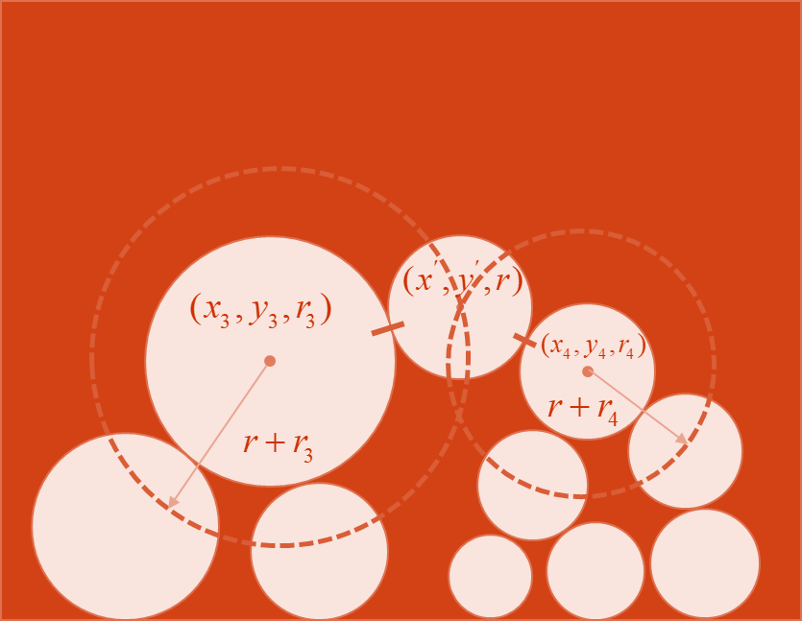}
    \caption{Position after collision}
    \label{fig:collision2}
\end{figure}

When the new particle stabilizes by resting on two pre-existing particles, whose coordinates are \((x_3, y_3, r_3)\) and \((x_4, y_4, r_4)\), respectively, the final position of the new particle \((x_{\text{new}}, y_{\text{new}})\) satisfies the following equations (see \textbf{Figure \ref{fig:collision2}}):

\begin{align}
    (x_{\text{new}} - x_3)^2 + (y_{\text{new}} - y_3)^2 &= (r + r_3)^2 \\
    (x_{\text{new}} - x_4)^2 + (y_{\text{new}} - y_4)^2 &= (r + r_4)^2
\end{align}

\begin{figure}[h]
    \centering
    \includegraphics[width=0.5\linewidth]{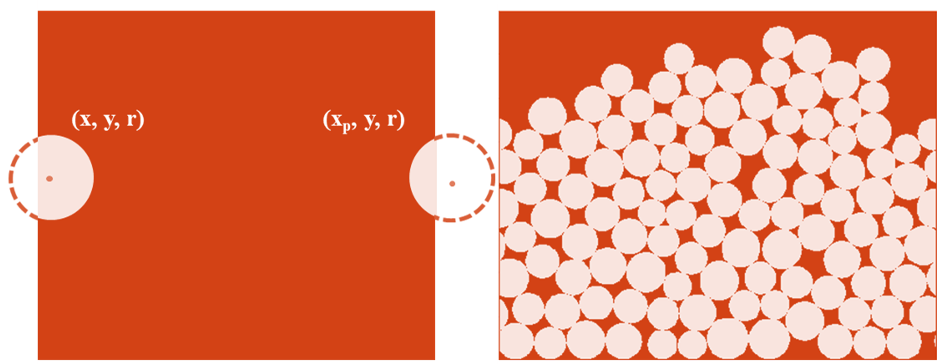}
    \caption{Periodic boundary treatment}
    \label{fig:period}
\end{figure}

To minimize boundary effects that might artificially increase the porosity near the edges of the simulation area, periodic boundary conditions are applied. This means that when a particle crosses the left boundary, it re-enters from the right, and vice versa. This boundary condition ensures that the particle packing reflects an infinite packing arrangement without artificial constraints caused by fixed boundaries. For example (\textbf{Figure \ref{fig:period}}), when a particle \( (x, y, r) \) is falling or rotating upon collision, if it exceeds the left boundary of the specified simulation domain, it re-enters the designated packing region from the right, with new coordinates \( (x_p, y, r) \), where: \(x_p = x + L\)
Similarly, if the particle exceeds the right boundary of the area, it re-enters from the left with coordinates \( (x_p, y, r) \), given by:
\(x_p = x - L\). 

The concept of periodic boundaries ensures that when particles generated in the simulation exceed the left or right boundaries of the packing region, they pass through and re-enter from the opposite side. This mechanism promotes a tightly packed arrangement of particles, eliminating the negative effects that traditional boundaries have on porosity calculations. Periodic boundaries can be viewed as enabling packing over an infinite range within a finite region. Although the packing region has a physical length of \( L \), periodic boundaries effectively extend the domain, approximating an infinitely long packing area. This approach broadens the scope of the simulation and enhances the reliability of porosity calculations.

\begin{figure}[h]
    \centering
    \includegraphics[width=0.5\linewidth]{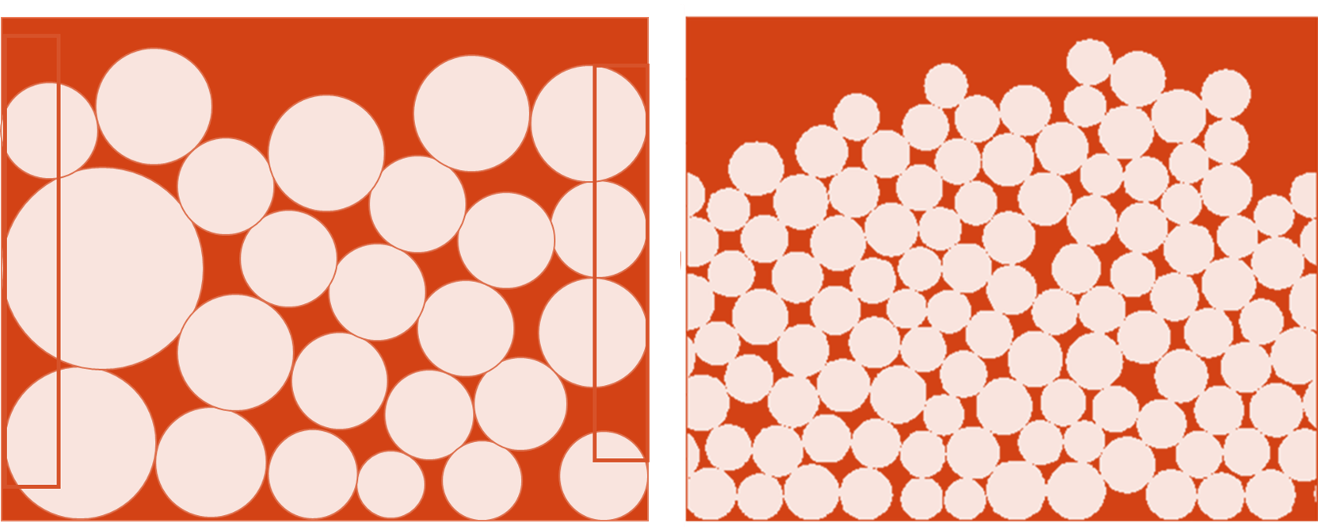}
    \caption{Boundary effect}
    \label{fig:boundary}
\end{figure}

Without periodic boundaries, particles near the edges of the packing region would not form a dense arrangement, leading to abnormally large porosity near the boundaries. As shown in \textbf{Figure \ref{fig:boundary}}, traditional boundaries result in artificially high porosity at the edges, distorting the overall porosity of the packed structure. This overestimation compromises the accuracy of the relationship between porosity and particle size distribution. By applying periodic boundaries, the issue of excessive edge porosity is avoided, ensuring a more accurate and realistic representation of the packed structure. Periodic boundaries, by extending the effective simulation area, improve the precision of porosity calculations and more closely represent real-world systems, where particles are able to pack densely without interference from artificial boundary effects.

\begin{figure}[h]
    \centering
    \includegraphics[width=0.3\linewidth]{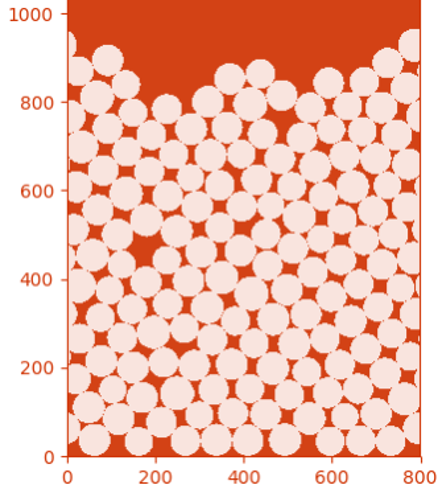}
    \caption{Packed particles}
    \label{fig:packed}
\end{figure}

Repeat the above steps to generate a series of particles. Suppose the designated height of the packed particle structure is \( h \). When a particle is generated such that its center's vertical coordinate exceeds the height \( h \), the particle generation process is completed, and no further particles are generated. Since each particle eventually settles either on top of previously deposited particles or directly on the ground, the resulting structure is a tightly packed two-dimensional particle assembly. For a visual representation, refer to \textbf{Figure \ref{fig:packed}}. For the generated two-dimensional particle assembly, the porosity calculation is straightforward.

\section{Application}

An example is provided below to demonstrate the application of the aforementioned simulation approach in correlating porosity with particle size distribution. By specifying different mean particle sizes \( \mu \) and standard deviations \( \sigma \), various combinations of \( \mu \) and \( \sigma \) are designed. The range of \( \mu \) is from 20 to 40 with an interval of 1, while \( \sigma \) ranges from 0 to \( \sigma_{\text{max}} \), also with an interval of 1. \(\sigma_{max} = \text{int}(\frac{\mu}{2}) - \text{int}(\frac{\mu}{10} - 2)\), where \( \text{int}(\cdot) \) represents the floor function, which returns the greatest integer less than or equal to the input. This results in a total of 235 combinations of \( (\mu, \sigma) \), thus generating \( N = 235 \) two-dimensional particle assemblies. The porosity \( \phi_i \) is calculated for each particle assembly, where \( i = 1, 2, 3, \dots, 235 \). 

The relative standard deviation \( \sigma / \mu \) is used as a parameter to describe the particle size distribution. It is calculated for each packing structure, where \( i = 1, 2, 3, \dots, N \). The primary reason for using the relative standard deviation \( \sigma / \mu \) to characterize the particle size distribution is that, generally, the mean particle size \( \mu \) and the standard deviation \( \sigma \) are the most common parameters for describing particle size distributions. However, qualitative analysis reveals that porosity does not change significantly with the mean particle size \( \mu \). In other words, simply altering the mean particle size (i.e., the size of the particles) does not affect porosity. On the other hand, porosity is highly sensitive to changes in the particle size standard deviation \( \sigma \), which reflects the degree of sorting of the particles. Porosity varies significantly with different levels of particle size sorting.

Nonetheless, the particle size standard deviation \( \sigma \) cannot be meaningfully compared without considering the mean particle size \( \mu \). For example, a collection of small particles with a large size variation may have a small mean particle size but a large relative standard deviation. In contrast, another set of particles that are much larger and more uniformly distributed may have a large mean particle size but a small relative standard deviation. In such cases, these two collections of particles cannot be meaningfully compared using only the relative standard deviation.

Therefore, the relative standard deviation must not be considered independently of the mean particle size when comparing particle size distributions. As such, this method proposes using the relative standard deviation \( \sigma / \mu \) as the key parameter for describing particle size distribution, which will be used in determining subsequent quantitative relationships.

\begin{figure}[h]
    \centering
    \includegraphics[width=0.5\linewidth]{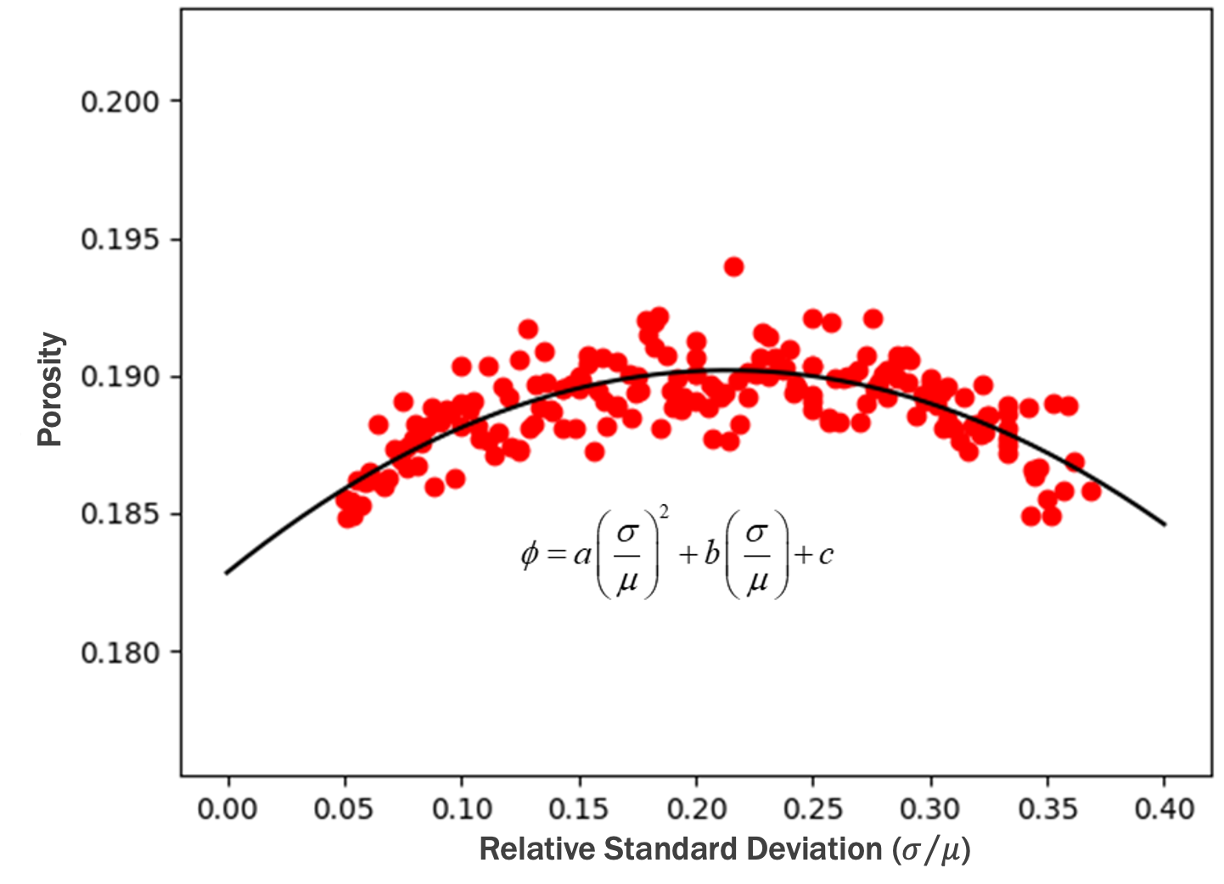}
    \caption{The correlation between porosity and relative standard deviation}
    \label{fig:correlation}
\end{figure}

For the generated samples, the porosities and the corresponding values of \( \sigma / \mu \) are plotted in \textbf{Figure \ref{fig:correlation}}. It can be observed that there is a parabolic relationship between porosity and the relative standard deviation. The relationship between them is defined by the following expression:

\[
\phi = a \left(\frac{\sigma}{\mu}\right)^2 + b \left(\frac{\sigma}{\mu}\right) + c
\]

The coefficients \( a \), \( b \), and \( c \) can be determined through quadratic regression. In this example, the values of \( a \), \( b \), and \( c \) are found to be \(-0.1611\), \(0.06882\), and \(0.1818\), respectively. This completes the qualitative and quantitative characterization of the relationship between porosity and particle size distribution.

It is evident that porosity is strongly correlated with the relative standard deviation. As the relative standard deviation increases, the particle size distribution becomes more uneven, with smaller particles blocking the pores created by larger particles, leading to lower porosity. Conversely, when the relative standard deviation decreases, the particles are better sorted, resulting in a tighter packing and similarly lower porosity. In this case, the quantitative relationship between porosity and the relative standard deviation follows a parabolic pattern, which can be determined through nonlinear regression analysis. This confirms the feasibility of the two-dimensional particle packing simulation method proposed in this study for characterizing the relationship between porosity and particle size distribution.

\section{Summary}

In this note, a simulation method is presented to explore the interplay between porosity and particle size distribution. The method generates particles with random coordinates and radii, mimicking the actual size distributions observed in rock formations. A set of rules is applied to simulate particle descent, collision, and stabilization within the packing domain. Periodic boundary conditions are implemented to mitigate edge effects and ensure more accurate porosity calculations. By systematically varying the mean particle size and standard deviation, multiple packed structures are generated and their respective porosities are computed. The resulting data are used to quantitatively and qualitatively assess the influence of particle size distribution on porosity. This simulation framework provides valuable insights into how different particle size distributions impact the packing and porosity of porous media, offering a robust tool for future studies in this domain.

\bibliographystyle{unsrt}
\bibliography{porosity-psd}

\end{document}